\documentclass[letterpaper, 10 pt, conference]{ieeeconf}
\IEEEoverridecommandlockouts
\usepackage{cite}
\usepackage{amsmath,amssymb,amsfonts,mathrsfs}
\usepackage{algorithm}
\usepackage{algorithmic}
\usepackage{graphicx}
\usepackage{hyperref}
\usepackage{textcomp}
\usepackage{xcolor}
\usepackage{comment}
\def\BibTeX{{\rm B\kern-.05em{\sc i\kern-.025em b}\kern-.08em
    T\kern-.1667em\lower.7ex\hbox{E}\kern-.125emX}}
    
\begin{document}
\title{\LARGE \bf Semi-Data-Driven Model Predictive Control: \\A Physics-Informed Data-Driven Control Approach}

\author{Sebastian Zieglmeier\textsuperscript{1}, Mathias Hudoba de Badyn\textsuperscript{1}, Narada D. Warakagoda\textsuperscript{1,2}, \\
Thomas R. Krogstad\textsuperscript{2} and Paal Engelstad\textsuperscript{1}
\thanks{The authors are with the Department of Technology Systems\textsuperscript{1} at the University of Oslo, Kjeller, Norway and the Norwegian Defence Research Establishment\textsuperscript{2}, Kjeller, Norway.}
\thanks{e-mails: \texttt{\{sebastiz, mathihud\}@uio.no}.}%
\thanks{Code available under: \href{https://github.com/SebsDevLab/SD-MPC.git}{https://github.com/SebsDevLab/SD-MPC.git}}
}

\maketitle

\thispagestyle{empty}
\pagestyle{empty}

\begin{abstract}
Data-enabled predictive control (DeePC) has emerged as a powerful technique to control complex systems without the need for extensive modeling efforts. However, relying solely on offline collected data trajectories to represent the system dynamics introduces certain drawbacks. Therefore, we present a novel semi-data-driven model predictive control (SD-MPC) framework that combines (limited) model information with DeePC to address a range of these drawbacks, including sensitivity to noisy data and a lack of robustness.  
In this work, we focus on the performance of DeePC in operating regimes not captured by the offline collected data trajectories and demonstrate how incorporating an underlying parametric model can counteract this issue.
SD-MPC exhibits equivalent closed-loop performance as DeePC for deterministic linear time-invariant systems. 
Simulations demonstrate the general control performance of the proposed SD-MPC for both a linear time-invariant system and a nonlinear system modeled as a linear parameter-varying system. These results provide numerical evidence of the enhanced robustness of SD-MPC over classical DeePC.
\end{abstract}

\section{Introduction}
\label{section:Introduction}
For many years, model predictive control (MPC) has been a widely adopted control strategy due to its ability to handle multivariable control problems with constraints \cite{Camacho_2007}. However, the performance of MPC is highly dependent on the accuracy of the underlying process model, which can be challenging to obtain, especially for complex or nonlinear systems \cite{Verheijen_2023}. Identifying the process of a system and modeling it accurately is often the most time-consuming step in developing an MPC \cite{Elokda_2021}.
To address this limitation, data-driven control has gained interest as an alternative to explicit parametric models, especially for complex systems \cite{Verheijen_2023, Elokda_2021}. 
One such data-driven control approach is data-enabled predictive control (DeePC; also called data-driven MPC), which uses a data-based system representation based on the fundamental lemma from behavioral theory to predict input-output trajectories directly \cite{Coulson_2019}. This reduces the modeling effort and can mitigate model errors and inaccuracies. A further key benefit of DeePC is the inheritance of stability and constraint satisfaction guarantees of conventional MPC \cite{Berberich_2020}.

Although DeePC offers numerous advantages, it also faces several limitations. Despite the parametric-free nature of DeePC, the performance is still dependent on the quantity and quality of the data. Insufficient data can lead to poor prediction accuracy, compromising the closed-loop performance \cite{Huang_2023}. Robust methods have been proposed to address this limitation in \cite{Huang_2023, Coulson_2022}. This is closely related to the high sensitivity to unseen operating points, as DeePC relies in general solely on offline data. If the collected data trajectories do not encompass all operating conditions, the performance of the controlled system can degrade significantly. Consequently, DeePC lacks robustness to certain inaccuracies compared to MPC, which uses physical insights to extrapolate. Existing robustness enhancements of DeePC \cite{Berberich_2020, Huang_2023, Coulson_2022, Coulson_2021} do not adequately address this issue of unseen operating points. 
Additionally, the Fundamental Lemma underlying DeePC is specific to linear time-invariant (LTI) systems, restricting its applicability to nonlinear or noisy systems \cite{Coulson_2019}. However, suitable regularization techniques have enabled DeePC to perform well on noisy and weakly nonlinear systems \cite{Elokda_2021, Coulson_2021, Berberich_2021}.
Finally, DeePC has limited adaptability, requiring specialized techniques for collecting persistently exciting data online \cite{Verheijen_2023, Berberich_2021}.

To address these limitations, we propose a semi-data-driven approach that integrates (potentially limited) prior model information with DeePC and leverages the strengths of both model-based and data-driven control. The core concept is to utilize the parametric model to capture the fundamental system dynamics, which are typically more straightforward to model, while employing the data-driven model to address the more complex residuals of the system.
The proposed semi-data-driven model predictive control (SD-MPC) combines a parametric and a data-driven model and is anticipated to demonstrate enhanced robustness to unseen operating points. SD-MPC should handle these inaccuracies more effectively, exhibiting characteristics more similar to the robust properties of MPC than the solely data-driven DeePC. Furthermore, the SD-MPC framework enables the incorporation of physical insights while still capitalizing on the advantages of reduced modeling effort provided by the data-driven components.

Our contributions are as follows: 
\begin{itemize}
\item We propose a semi-data-driven model predictive control framework that combines DeePC with (potentially limited) knowledge of a model.
\item We demonstrate the effectiveness of the proposed approach through numerical simulations for an LTI system and a linear parameter-varying (LPV) system.
\item We demonstrate how SD-MPC can enhance the robustness of data-driven control approaches to inaccuracies.
\end{itemize}

Further hybrid data-driven approaches are \cite{Watson_2025, Lazar_2023}. In \cite{Watson_2025}, the authors combine a parametric model with a data-driven approach, though with a different problem formulation. 
In \cite{Lazar_2023}, the authors combined subspace and Hankel predictors to reduce the online computational complexity. Both approaches differ clearly from the method presented in this work. 

The remainder of the paper is organized as follows: Section \ref{section:MPC} and Section \ref{section:DeePC} provide an overview of MPC and DeePC to introduce the notation, respectively. Section \ref{section:Semi} presents the proposed SD-MPC approach and discusses its properties with the according theoretical results in \ref{section:Theory}. Section \ref{section:results} showcases the numerical results, and Section \ref{section:Conclusion} concludes the paper.

\section{Model Predictive Control}
\label{section:MPC}
Consider the discrete LTI system:
\begin{equation}
\begin{aligned}
& x_{k+1}=A x_k+B u_k \\
& y_k=C x_k+D u_k,
\end{aligned}
\label{eq:LTI system}
\end{equation}
where $x_k \in \mathbb{R}^{n_x}$ is the state vector, $u_k \in \mathbb{R}^{n_u}$ is the input vector, and $y_k \in \mathbb{R}^{n_y}$ is the output vector at the discrete time step $k$. The matrices $A \in \mathbb{R}^{n_x \times n_x}$, $B \in \mathbb{R}^{n_x \times n_u}$, $C \in \mathbb{R}^{n_y \times n_x}$, $D \in \mathbb{R}^{n_y \times n_u}$ describe the system dynamics.

Given a desired reference trajectory $\{r_k \in \mathbb{R}^{n_y}\}_{k=0}^{N-1}$, we can formulate the receding-horizon MPC to solve the following trajectory tracking problem at each discrete time step $k$ \cite{Camacho_2007}:
\begin{equation}
\resizebox{.9\columnwidth}{!}{$\begin{array}{cl}
\underset{u, x, y}{\operatorname{minimize}} & \sum_{k=0}^{N-1}\left(\left\|y_k-r_{k}\right\|_Q^2+\left\|u_k\right\|_R^2\right) \\
\text { subject to } & x_{k+1}=A x_k+B u_k, \forall k \in\{0, \ldots, N-1\}, \\
& y_k=C x_k+D u_k, \forall k \in\{0, \ldots, N-1\}, \\
& x_0=\hat{x}(t), \\
& u_k \in \mathcal{U}, \forall k \in\{0, \ldots, N-1\}, \\
& y_k \in \mathcal{Y}, \forall k \in\{0, \ldots, N-1\},
\end{array}$}
\label{eq:MPC}
\end{equation}
where N is the prediction horizon. Constraints can be applied with the sets $\mathcal{U} \subseteq \mathbb{R}^{n_u}$and $\mathcal{Y} \subseteq \mathbb{R}^{n_y}$ on the input and output, respectively. Via the objective function with the quadratic norms weighted by the output cost matrix $Q \in \mathbb{R}^{n_y \times n_y}$ and the input cost matrix $R \in \mathbb{R}^{n_u \times n_u}$, the MPC aims to compute the optimal input sequence $u$ such that the predicted output tracks the reference trajectory while minimizing the control effort and satisfying constraints. In every iteration of the control loop, the first value $u_0$ of the computed input sequence $u$ is applied to the system as the control input. Either via measuring the state or by state estimation, the new state $\hat{x}(t)$ is obtained, and the optimization problem is solved again at the subsequent time step. To track the trajectory $r_k$, the controller must have an accurate model of the system. This constitutes an important limitation of classical MPC, as obtaining such an accurate model can be challenging, especially for complex systems.

\section{Data-Driven Predictive Control}
\label{section:DeePC}
DeePC provides an alternative solution by directly using input-output measurement data, thereby avoiding the need for an explicit model of the system. 

\subsection{Preliminaries}

First, we briefly review the key definitions and lemmas introduced in \cite{Coulson_2019} relevant for the subsequent work.
In DeePC a sequence $u = col(u_1, …, u_T) \in \mathbb{R}^{Tn_u}$ can be represented using a Hankel matrix $\mathscr{H}_L(u)$ with the column length ${L}$ as follows: 
\begin{equation}
\mathscr{H}_L(u):=\left(\begin{array}{ccc}
u_1 & \cdots & u_{T-L+1} \\
\vdots & \ddots & \vdots \\
u_L & \cdots & u_T
\end{array}\right)
\label{eq:Hankelmatrix}
\end{equation}
\textbf{Definition 1:} (\cite{Coulson_2019}, Definition 4.4)
\label{definition1}
Let $L,T \in \mathbb{Z}_{>0}$ such that $T \geq L$. The sequence ${u}$ is persistently exciting of order $L$ if the corresponding Hankel matrix $\mathscr{H}_L(u)$ has full row rank.

Furthermore, DeePC is based on Willems' fundamental lemma \cite{Willems_2004}, which states that the entire set of trajectories for an LTI system can be reproduced from a single persistently exciting trajectory of sufficient length~\cite{Coulson_2019}. In other words, one persistently exciting input and output trajectory captures the complete behavior of a system. 

\textbf{Theorem 1:} (\cite{Willems_2004}, Theorem 1, \cite{Coulson_2019}, Lemma 4.2)
Consider a controllable system $\mathscr{B}$, which is linear, time-invariant, and complete (see \cite{Coulson_2019}, Definition 4.2). Let $T, t \in \mathbb{Z}_{>0}$ and  $w = col(u; y) \in \mathscr{B}_T$, where $\mathscr{B}_T$ is the restricted behavior, i.e. a set of trajectories truncated to a window of length $T$. With a persistently exciting input ${u}$ of order $t+n(\mathscr{B})$, where $n$ denotes the order of $\mathscr{B}$, it follows that $colspan(\mathscr{H}_t(w)) = \mathscr{B}_t$. 
Therefore, the system can be fully represented as a linear combination of the columns in the Hankel matrix $\mathscr{H}_t(w)$.

\subsection{DeePC Formulation}

Let $T, T_{ini}, N \in \mathbb{Z}_{>0}$ with ${T}$ being sufficiently large according to Definition 1. By applying the persistently exciting sequence $u^d = col(u^d_0, …, u^d_{T-1}) \in \mathbb{R}^{Tn_u}$ to the controllable system $\mathscr{B}$ we receive the corresponding outputs $y^d = col(y^d_0, …, y^d_{T-1}) \in \mathbb{R}^{Tn_y}$. The collected data trajectories $u^d$ and $y^d$ are split up in past data and future data \cite{Coulson_2019}: 
\begin{equation}
\binom{U_{{p}}}{U_{{f}}}:=\mathscr{H}_{T_{\text {ini }}+N}\left(u^{{d}}\right),  \binom{Y_{{p}}}{Y_{{f}}}:=\mathscr{H}_{T_{\text {ini }}+N}\left(y^{{d}}\right),
\label{eq:Hankelmatrices_split}
\end{equation}
where the past matrices $U_p$ and $Y_p$ hold the first $T_{ini}$ number of rows and the future matrices $U_f$ and $Y_f$ hold the last ${N}$ number of rows of the Hankel matrices $\mathscr{H}_{T_{ini}+N}$. The explicit model required for traditional MPC in \eqref{eq:MPC} can be substituted with a data-driven representation leveraging the Hankel matrices with the past and future data \cite{Coulson_2019}.
\begin{equation}
\resizebox{.75\columnwidth}{!}{$\begin{array}{ll}
\underset{g, u, y, \sigma_y}{\operatorname{minimize}} & \sum_{k=0}^{N-1}\left(\left\|y_k-r_{k}\right\|_Q^2+\left\|u_k\right\|_R^2\right) \\
& +\lambda_g\|g\|_1+\lambda_y\left\|\sigma_y\right\|_1 \\
\text { subject to } & \left(\begin{array}{c}
U_{{p}} \\
Y_{{p}} \\
U_{{f}} \\
Y_{{f}}
\end{array}\right) g=\left(\begin{array}{c}
u_{ {ini}} \\
y_{ {ini}} \\
u \\
y
\end{array}\right)+\left(\begin{array}{c}
0 \\
\sigma_y \\
0 \\
0
\end{array}\right) \\
& u_k \in \mathcal{U}, \forall k \in\{0, \ldots, N-1\} \\
& y_k \in \mathcal{Y}, \forall k \in\{0, \ldots, N-1\}
\end{array}$}
\label{eq:DeePC}
\end{equation}
Here, $u_{ini}$ and $y_{ini}$ represent the last $T_{ini}$ inputs and outputs of the system, respectively, while ${u}$ and ${y}$ are the future inputs and outputs to be optimized. Building upon Theorem 1, the decision vector ${g} \in \mathbb{R}^{T-T_{ini}-N+1}$ linearly combines the collected data with the current system's data. 
The norms $||{g}||$ and $||\sigma_y||$, where $\sigma_y \in \mathbb{R}^{T_{ini}n_y}$, along with the corresponding regularization weights $\lambda_g, \lambda_y \in \mathbb{R}_{>0}$ were introduced in \cite{Coulson_2019} to enhance the robustness of DeePC against various system complexities, such as noise, nonlinearities, and time delays, which may arise when controlling systems that deviate from the assumption of linearity and time-invariance.

\section{Semi-Data-Driven Model Predictive Control}
\label{section:Semi}
DeePC's performance degrades when the controlled system deviates from the collected data, particularly within unseen operating points. 
This occurs if the stored data in the Hankel matrices $U_p, Y_p, U_f, Y_f$, which is collected offline, is unrepresentative of online system dynamics.

\begin{figure}
    \centering
    \includegraphics[width=0.5\linewidth]{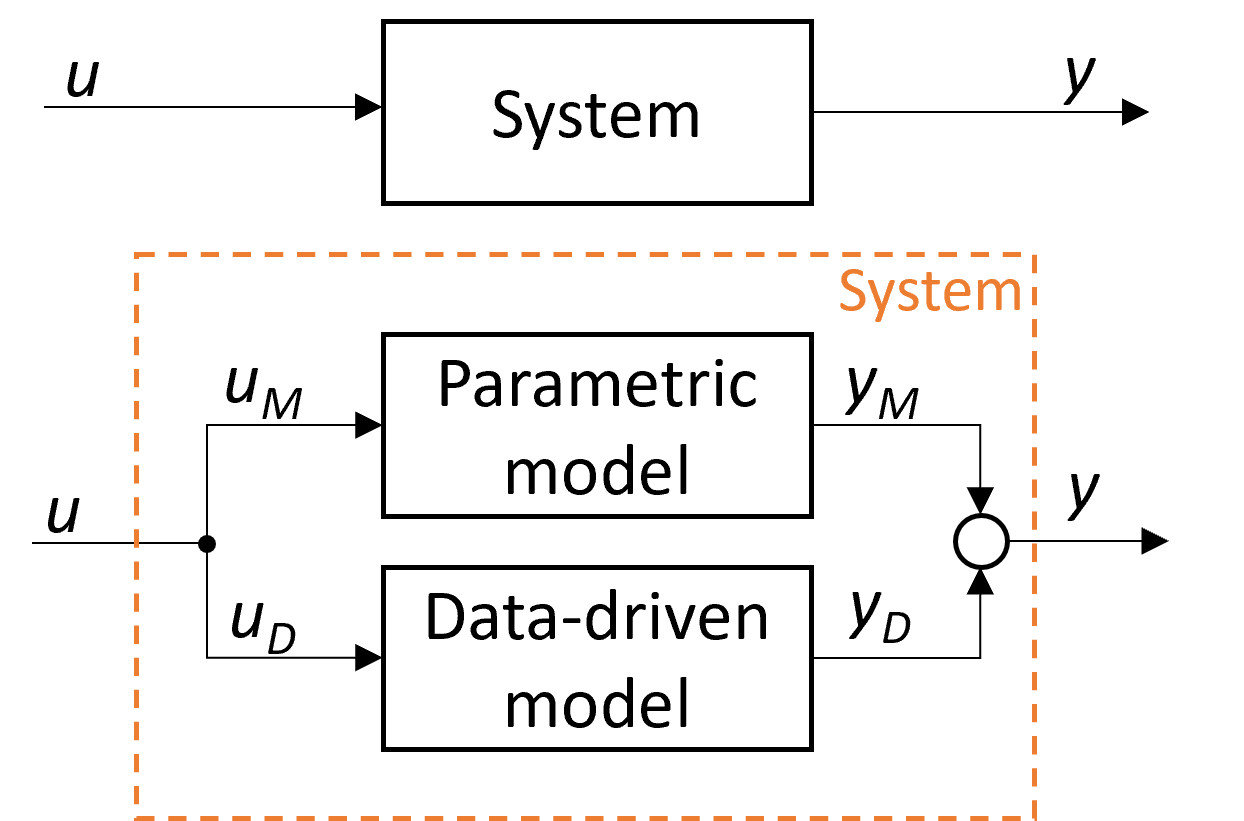}
    \caption{Representation of the system with two parallel subsystems leveraging a parametric and a data-driven model. }
    \label{fig:block_diagramm}
\end{figure}

To address this issue, we propose the semi-data-driven model predictive control approach, which combines the data-driven DeePC with a model-based MPC to leverage the advantages of both. 
Due to incomplete system knowledge, a parametric model estimates the fundamental behavior, while DeePC captures residual dynamics, maintaining a low modeling effort. 
Therefore, the SD-MPC framework models the system as a composition of two parallel subsystems, a parametric and a data-driven model, as illustrated in Figure~\ref{fig:block_diagramm}. To preserve the simplicity of the initial presentation of SD-MPC, more complex subsystem compositions are left for future investigation. Similarly, SD-MPC is developed and applied to an LTI system in the current work, enabling proofs and a comparative analysis against the DeePC method.

\subsection{SD-MPC Formulation}

The SD-MPC approach partitions the system into two components: the fundamental behavior and the residual behavior. The fundamental behavior represents the dominant and well-characterized system dynamics, which are captured by a parametric LTI model. In contrast, the residual behavior accounts for the remaining uncertainties and unmodeled dynamics not covered by the parametric model.
Although existing DeePC-based approaches \cite{Verheijen_2023, Elokda_2021, Coulson_2019} often assume no prior knowledge of the system dynamics, we argue that some understanding of fundamental system behavior generally exists. 
The available prior knowledge regarding the fundamental system dynamics can be utilized to formulate a parametric model, marked with the subscript $M$ in \eqref{eq:Semi}, which describes the main system behavior. Additionally, we assume that full state and output measurements of the controlled system are available.
Applying a persistently exciting input sequence $u^d$ in parallel to the parametric model and the actual system yields the outputs $y^d_M$ and ${y}^d$, respectively.

In \cite{Willems_2007}, Willems defined the fundamentals of the behavioral approach for interconnected systems via tearing, zooming, and linking. While Willems introduces these methods for modeling the interaction of real existing subsystems inside a whole system, the provided methods also hold for our imaginary non-existing subsystems, the estimated main behavior, and the resulting residual behavior. Thus, we can formulate the following definition and lemma as the basis of SD-MPC. 

\textbf{Definition 2:} Two parallel subsystems $\mathscr{B}_1$ and $\mathscr{B}_2$ represent a system $\mathscr{B}$, provided they share the same input $u^d$ $\in \mathbb{R}^{Tn_u}$ and their respective outputs $y^d_1\in \mathbb{R}^{Tn_y}$ and $y^d_2 \in \mathbb{R}^{Tn_y}$ add up to the overall output $y^d \in \mathbb{R}^{Tn_y}$, i.e. $y^d = y^d_1 + y^d_2$.

\textbf{Lemma 1:} Consider a controllable system $\mathscr{B}$ , which is linear, time-invariant and complete (see \cite{Coulson_2019}, Definition 4.2). Let $T \in \mathbb{Z}_{>0}$, $\mathscr{B}_1$ and $\mathscr{B}_2$ be the two parallel subsystems of the overall system $\mathscr{B}$ as per Definition 2.  Suppose $u^d$ $\in \mathbb{R}^{Tn_u}$ is a persistently exciting input applied to $\mathscr{B}, \mathscr{B}_1$ and $\mathscr{B}_2$ and $y^d = y^d_1 + y^d_2$, where $y^d$, $y^d_1$ and $y^d_2 \in \mathbb{R}^{Tn_y}$ represent the future outputs of $\mathscr{B}$, $\mathscr{B}_1$ and $\mathscr{B}_2$, respectively. If $\mathscr{B}_1$ is linear, time-invariant, and complete, $\mathscr{B}_2$ is also linear, time-invariant, and complete.

In other words, with Definition 2 and Lemma 1, a system and its behavior can be decomposed into multiple parallel subsystems and their sub-behaviors. 
By application of Definition 2, we receive the residual trajectory ${y}^d_D = {y}^d - {y}^d_M$, representing the difference between the actual system behavior and the fundamental behavior captured by the parametric model for a given input $u^d=u_M^d=u_D^d$. Accordingly, as a further implication of Lemma 1, if the parametric LTI model does not accurately represent the LTI system, the residual behavior can also be modeled as a separate parallel LTI model. Therefore, the overall system behavior is the combination of the parametric model behavior and the data-driven residual behavior.
Consistent with DeePC, we store the sequences $u^d$ and $y_D^d$ in the Hankel matrices $U_p, U_f$ and $Y_{p,D}, Y_{f,D}$, respectively.
With the residual behavior stored (subscript $D$) in the Hankel matrices, and the fundamental behavior captured by $A_M \in \mathbb{R}^{n_x \times n_x}$, $B_M \in \mathbb{R}^{n_x \times n_u}$, $C_M \in \mathbb{R}^{n_y \times n_x}$ and $D_M \in \mathbb{R}^{n_y \times n_u}$ we can now formulate the SD-MPC problem in \eqref{eq:Semi}.
\begin{equation}
\resizebox{.85\columnwidth}{!}{$
\begin{array}{cl}
\underset{u, x, y, g, \sigma_y}{\operatorname{minimize}} & \sum_{k=0}^{N-1}\left(\left\|y_k-r_{k}\right\|_Q^2+\left\|u_k\right\|_R^2\right) \\
&+\lambda_g\|g\|_1+\lambda_y\left\|\sigma_y\right\|^2 \\
\text { subject to } & x_{k+1,M}=A_M x_{k,M}+B_M u_k, \forall k \in\{0, \ldots, N-1\}, \\
& y_{k,M}=C_M x_{k,M}+D_M u_k, \forall k \in\{0, \ldots, N-1\}, \\
& x_{0,M}=\hat{x}(t), \\
&\left(\begin{array}{c}
U_{{p}} \\
Y_{{p,D}} \\
U_{{f}} \\
Y_{{f,D}}
\end{array}\right) g=\left(\begin{array}{c}
u_{ {ini}} \\
y_{ {ini,D}} \\
u \\
y_D
\end{array}\right)+\left(\begin{array}{c}
0 \\
\sigma_y \\
0 \\
0
\end{array}\right), \\
& y_k = y_{k,M} + y_{k,D}, \\
& u_k \in \mathcal{U}, \forall k \in\{0, \ldots, N-1\}, \\
& y_k \in \mathcal{Y}, \forall k \in\{0, \ldots, N-1\},
\end{array}$}
\label{eq:Semi}
\end{equation}
As stated in Lemma 1, the overall system output ${y}$ is the sum of the output ${y}_M$ from the parametric model and the output ${y}_D$ from the data-driven model. This relationship is incorporated into the constraints, enabling tracking of the reference trajectory with the combined output in the objective function. As Lemma 1 indicates further, the parallel interconnection of the subsystems implies that both the parametric and data-driven models utilize the same control input ${u}$.
Compared to standard DeePC, where the Hankel matrices cover the whole behavior of the system, in SD-MPC, the Hankel matrices only cover the residual behavior. As we defined parallel interconnection of the subsystems, the initial input vector $u_{{ini}}$ can be filled with the last $T_{{ini}}$ inputs ${u}$ applied to the overall system. In contrast, the initial output vector $y_{{ini,D}} \in \mathbb{R}^{T_{ini}n_y}$ must represent the last $T_{{ini}}$ outputs of the residual behavior, rather than the outputs of the entire system. The newest value of the initial output vector $y_{ini,D}$ can be computed by subtracting the output generated by the parametric model from the most recent measured output of the complete system, therefore determining the impact of the control due to the data-driven model.
\begin{equation}
    y_{ini,D}(t) = y(t) - y_M(t)
    \label{eq:y_ini}
\end{equation}
Additionally, we present two distinct strategies for updating the state $x_{0,M}$ in the SD-MPC approach described in \eqref{eq:Semi}. The general SD-MPC method utilizes the parametric model to propagate its own state $x_{0,M}$ forward into the next iteration, reflecting the parametric model's state response to the control input $u_0$. This feedback strategy originates from the data collection process outlined in Lemma 1, where the data trajectories of the system and the parametrized model evolve independently over time.

The robust semi-data-driven model predictive control (rSD-MPC) method utilizes the current system state as feedback to update the parametric model's state at each iteration of the control loop. Updating the parametric model's state with the system's current state at each iteration can be necessary to prevent potential divergence and misalignment between the system state and the parametric model state over time. This state update approach helps especially to robustify the control against uncertainties in the system. Consequently, the rSD-MPC method requires updating the parametric model's state with the current state of the system at each iteration of the data collection process. This approach ensures that the collected data for the data-driven component represents the required residual behavior within the control process. This leads to the following algorithms \ref{alg:3} and \ref{alg:2} for the data collection and control process, where general SD-MPC involves steps 3.1 and 5.1, while rSD-MPC utilizes steps 3.2 and 5.2. 
The remainder of \eqref{eq:Semi} is consistent with standard DeePC detailed previously, sharing the same purpose and definitions.
\begin{algorithm}
\caption{Data collection process for SD-MPC/rSD-MPC}
\label{alg:3}
\begin{algorithmic}[1] 
    \STATE Generate an input $u_k^d \in \mathcal{U}$ satisfying Definition 1 and apply it to the system and record $y_k^d$, $x_k^d$, and $x_{k+1}^d$.     
    \STATE Calculate the output $y_{k,M}^d$ of the parametric model: $y_{k,M}^d = C_M x_{k,M}^d+D_M u_k^d$
    \STATE Update $x_{k,M}^d$:
    \begin{itemize} 
       \item[3.1] SD-MPC: No update of $x_{k,M}^d$ needed.
       \item[3.2] rSD-MPC: Synchronize the model's state vector $x_{k,M}^d$ with the state of the system $x_k^d$: $x_{k,M}^d \leftarrow x_k^d$ 
    \end{itemize} 
    \STATE Calculate the next state $x_{k+1,M}^d$ of the parametric model for the subsequent iteration: $x_{k+1,M}^d = A_M x_k^d + B_M u_k^d$ 
    \vspace{-\baselineskip}
    \STATE Return to step 1 until $k = T_d$.
    \STATE Calculate the data trajectory $y_{D}^d = y^d - y_M^d$ and $u_{D}^d = u^d$. 
\end{algorithmic}
\end{algorithm}
\begin{algorithm}
\caption{SD-MPC/rSD-MPC process}
\label{alg:2}
\begin{algorithmic}[1] 
    \STATE Solve \eqref{eq:Semi} for the optimal control input $u$ and extract the estimated output of the parametrized model $y_M$. 
    \STATE Apply the first control input $u_1$ to the system. 
    \STATE Measure the system's output $y$ and states $x_k$ and $x_{k+1}$. 
    \STATE Calculate the last value of $y_{ini,D}$ using \eqref{eq:y_ini} and update $u_{ini}$ and $y_{ini,D}$ to the $T_{ini}$ most recent values. 
    \STATE Update the state vector $x_{0,M}$ of the parametrized model
    \begin{itemize}
       \item[5.1] SD-MPC: with $x_{k+1,M}$.
       \item[5.2] rSD-MPC: with $x_{k+1}$.
   \end{itemize} 
    \STATE Return to 1
\end{algorithmic}
\end{algorithm}
\section{Theoretical Results}
\label{section:Theory}
\textbf{Proof of Lemma 1:}
Let the system be represented by its observability matrix $\mathscr{O}_{T_d}$ and its lower triangular Toeplitz matrix $\mathscr{T}_{T_d}$ with $T_{d}$ being the number of collected data-points and therefore the horizon of the $\mathscr{O}$-$\mathscr{T}$-representation: 
\begin{equation}
\begin{aligned}
y&=\mathscr{O}_{T_d} x_{0}+\mathscr{T}_{T_d} u \\
&=\resizebox{.7\columnwidth}{!}{$ \begin{bmatrix} C \\ CA  \\ \vdots \\ CA^{T_d-1} \end{bmatrix} x_0 + \left[\begin{array}{cccc}
D & 0 & \cdots & 0 \\
C B & D & \cdots & 0 \\
\vdots & \ddots & \ddots & \vdots \\
C A^{T_d-2} B & \cdots & C B & D
\end{array}\right] u$}
\label{eq:system_representation_general}
\end{aligned}
\end{equation}
Let the parametric model equivalently transformed into the $\mathscr{O}$-$\mathscr{T}$-representation, with its observability matrix $\mathscr{O}_{T_d,M}$ and the lower triangular Toeplitz matrix $\mathscr{T}_{T_d,M}$. Then by applying the introduced relation $y^d_D = y^d - y_M^d$, resulting out of Definition 2, we get the following $\mathscr{O}$-$\mathscr{T}$-representation:
\begin{equation}
\begin{aligned}
y_D =& y - y_M \\
=&\mathscr{O}_{T_d} x_{0}+\mathscr{T}_{T_d} u -\mathscr{O}_{T_d,M} x_{0,M}-\mathscr{T}_{T_d,M} u \\
\end{aligned}
\label{eq:Proof_lemma1_1}
\end{equation}
By assumption of the same initial state $x_0=x_{0,M}=x_{0,D}$ in the data-collection process and due to the shared input $u$ the data-driven subsystem can be represented as:
\begin{equation}
    y_D=\mathscr{O}_{T_d,D} x_{0}+\mathscr{T}_{T_d,D} u, 
    \label{eq:Proof_lemma1_2}
\end{equation}
with 
\begin{equation}
\begin{aligned}
&\mathscr{O}_{T_d,D} = \mathscr{O}_{T_d} -\mathscr{O}_{T_d,M} \\
&\mathscr{O}_{T_d,D} = \resizebox{.6\columnwidth}{!}{$\begin{bmatrix} C_D \\ C_DA_D \\ C_DA_D^2 \\ \vdots \\ C_DA_D^{T_d-1} \end{bmatrix}  = \begin{bmatrix} C-C_M \\ CA-C_MA_M \\ CA^2-C_MA_M^2 \\ \vdots \\ CA^{T_d-1}-C_MA_M^{T_d-1} \end{bmatrix}$}
\label{eq:data_observability_robust}
\end{aligned}
\end{equation}
and
\begin{equation}
\begin{aligned}
&\mathscr{T}_{T_d,D}=\mathscr{T}_{T_d}-\mathscr{T}_{T_d,M}  \\
&=\resizebox{.58\columnwidth}{!}{$\left[\begin{array}{cccc}
D_D & 0 & \cdots & 0 \\
C_D B_D & D_D & \cdots & 0 \\
\vdots & \ddots & \ddots & \vdots \\
C_D A_D^{T_d-2} B_D & \cdots & C_D B_D & D_D
\end{array}\right]$} \\
&=\resizebox{.83\columnwidth}{!}{$\left[\begin{array}{cccc}
D-D_M & 0 & \cdots & 0 \\
CB-C_M B_M & D-D_M & \cdots & 0 \\
\vdots & \ddots & \ddots & \vdots \\
C A^{T_d-2} B-C_M A_M^{T_d-2} B_M & \cdots &  & D-D_M
\end{array}\right]$}
\label{eq:data_toeplitz_robust}
\end{aligned}
\end{equation}
With this $\mathscr{O}$-$\mathscr{T}$-representation, we can conclude the proof that the second parallel subsystem is also an LTI system and therefore the residual behavior can be represented with the data-driven approach introduced in \cite{Coulson_2019}.
 
\textbf{Proposition 1 - Exact representation: }
Let the residual data be collected as described in Lemma 1 for a deterministic LTI system and assume no regularization ($\sigma_y=0$). By combining the output of the parametric and data-driven model, $y_M$ and $y_D$, respectively, in the SD-MPC \eqref{eq:Semi} with regard to their shared input $u$, the output of the system can be accurately predicted $y_{pred} = y_{pred,M} + y_{pred,D} = y$.

\textbf{Proof:}
As discussed previously, the parametric and data-driven model can be transferred to the $\mathscr{O}$-$\mathscr{T}$-representation, where the time horizon of the representation shifts to the prediction horizon $N$, and the collected data \eqref{eq:Proof_lemma1_1} captures the residual behavior as follows: 
\begin{equation}
y_{pred,M}=\mathscr{O}_{N,M} x_{0,M}+\mathscr{T}_{N,M} u .
\label{eq:proof_model_representation_robust}
\end{equation}
\begin{equation}
    y_{pred,D}=\mathscr{O}_{N} x_{0}+\mathscr{T}_{N} u -\mathscr{O}_{N,M} x_{0,M}-\mathscr{T}_{N,M} u
\end{equation}
For the overall predicted output in the general SD-MPC framework follows \eqref{eq:proof_exact_representation}.
\begin{equation}
\begin{aligned}
    y_{pred} =& y_{pred,M} + y_{pred,D} \\
    =&\mathscr{O}_{N,M} x_{0,M}+\mathscr{T}_{N,M} u + \mathscr{O}_{N} x_{0}+\mathscr{T}_{N} u \\
    &-\mathscr{O}_{N,M} x_{0,M}-\mathscr{T}_{N,M} u \\
    =& \mathscr{O}_{N} x_{0}+\mathscr{T}_{N} u
    \label{eq:proof_exact_representation}
\end{aligned}
\end{equation}
This predicted output aligns precisely with the output of the deterministic LTI system within the prediction horizon $N$, if the calculated $u$ is applied to the system.

\textbf{Proposition 2 - Recursive Feasibility:}
Given a deterministic LTI system described by \eqref{eq:LTI system} with the input and output constraint sets $\mathcal{U}$ and $\mathcal{Y}$, the general SD-MPC formulation is recursive feasible under the following assumptions:
\begin{itemize}
    \item The SD-MPC problem is initially feasible at $t=0$.
    \item There exist control actions within $\mathcal{U}$, which are able to sustain the system's output inside a given output constraint set $\mathcal{Y}$ and output terminal constraint set $\mathcal{Y}_T$ (comparatively, \cite{Berberich_2020}, Theorem 2). 
    \item The data collection for the residual data-driven model was executed, satisfying Lemma 1. 
\end{itemize}
Under these assumptions, if the SD-MPC optimization problem is feasible at time $k$, then it remains feasible at time $k+1$, ensuring recursive feasibility throughout the control horizon.

\textbf{Proof: (comparable to \cite{Mayne_2000} 3.3) } 
Let $Y_k$ denote, for each discrete time-step k, the set of outputs $y$, controllable by a feasible control sequence to the terminal constraint set $\mathcal{Y}_T$ in N steps or less. 
Furthermore, the control, output and terminal constraints, i.e. $u_k \in \mathcal{U}, \forall k \in \{0,1,\ldots, N-1\}$, $y_k \in \mathcal{Y}, \forall k \in \{0,1,\ldots, N-1\}$ and $y_N \in \mathcal{Y}_T$ are satisfied by a feasible control sequence $u=( u_0, u_1, \ldots, u_{N-1})$ at the discrete time step $k=0$. 
Therefore, the set of outputs that can be steered by \eqref{eq:Semi} is $Y_N$  with the prediction horizon $N$.
Suppose, that $y\in Y_N$, and the determined control sequence $u$ is the solution of \eqref{eq:Semi} at $k=0$ and let $y=(y_0, y_1, \ldots, y_{N-1})$ be then the optimal predicted output trajectory. The first control input $u_0$ of $u=( u_0, u_1, \ldots, u_{N-1})$ steers the initial output $y$ to its successor output $y^+$. 
To obtain a feasible control sequence for the next time step $k=1$, we append an additional control input $\bar u$ to the existing sequence, resulting in the extended sequence $u^+=(u_1, \ldots, u_{N-1}, \bar u)$. 
This control sequence is feasible, if $\bar u \in \mathcal{U}$, $\mathcal{Y}_T \subset \mathcal{Y}$, $\mathcal{Y}_T$ is positively control invariant and $\bar u$ steers the output $y^+=(y_1, \ldots, y_{N-1}, \bar y)$ to $\bar y \in \mathcal{Y}_T$.
As the control problem is feasible for time steps $k=0$ and $k=1$, it is feasible for all subsequent time steps, which concludes the proof of recursive feasibility.

\textbf{Proposition 3 - Equivalence of SD-MPC and MPC:}  SD-MPC is identical to MPC, if and only if the parametric model exactly represents the system.

\textbf{Proof:} Let the assumed parametric model be an exact representation of the system, with $A_M = A, B_M = B, C_M = C, D_M = D$. Following Lemma 1, the residual trajectory results in $y_D^d = 0$ for a deterministic LTI system. Hence, the optimization problem in \eqref{eq:Semi} simplifies to the standard MPC optimization problem in \eqref{eq:MPC}, using only the parametric model. 

\textbf{Proposition 4 - Equivalence of SD-MPC and DeePC:} SD-MPC is identical to DeePC, if and only if the parametric model represents no information about the system. 

\textbf{Proof:} Let the parametric model parameters be $A_M = 0, B_M = 0, C_M = 0, D_M = 0$. Following Lemma 1, the system output is fully described by the residual trajectory $y_D^d = y^d$, resulting in the optimization problem in \eqref{eq:Semi} simplifying to the DeePC optimization problem in \eqref{eq:DeePC}. 

These propositions demonstrate that the SD-MPC framework encompasses both MPC and DeePC as special cases. As the parametric model's accuracy increases, the semi-data-driven approach resembles classical MPC more closely. Conversely, as the parametric model deviates further from the true system, the semi-data-driven approach increasingly resembles DeePC. Therefore, the closer the parametric model is to the system, the more robust the semi-data-driven approach becomes to unseen operating points, without the need for unrealistic, extensive modeling efforts, as the residual data trajectories cover the differences. Robustness guarantees and theoretical proofs for rSD-MPC are left to future work. 

\section{Numerical Results}
\label{section:results}
To assess the general performance and robustness of the proposed SD-MPC approach, we conduct the following simulation study: We consider a cascaded two-tank system, where the control objective is to track the water level in the second tank by controlling the inflow into the first tank. The derivation of the state space model and specific parameter values for this system are presented in \cite{Teppa_2025}. Furthermore, for the following experiments, we assume noise-free measurements. The nonlinear two-tank system is formulated as an LPV system, as detailed below.
\begin{equation}
\begin{aligned}
&\resizebox{.85\columnwidth}{!}{$\dot{x}(t)=\left[\begin{array}{cc}
-0.904 \theta_1(t) & 0 \\
0.904 \theta_1(t) & -0.508 \theta_2(t)
\end{array}\right] x(t)+\left[\begin{array}{c}
0.258 \\
0
\end{array}\right] u(t)$}\\
&y(t)=\left[\begin{array}{ll}
0 & 1
\end{array}\right] x(t)
\end{aligned}
\label{eq:LPV system}
\end{equation}
where $\theta(t)$ presents the dependency of the water level in the tank: 
\begin{equation}
\boldsymbol{\theta}(t)=\left[\begin{array}{ll}
\theta_1(t) & \theta_2(t)
\end{array}\right]^T=\left[\begin{array}{ll}
1 / \sqrt{x_1} & 1 / \sqrt{x_2}
\end{array}\right]^T
\label{eq:theta}
\end{equation}
\subsection{LTI system}
\label{subsec:LTI}
To initially evaluate the general performance of the SD-MPC method, we compare its closed-loop behavior on standard LTI systems against the results obtained from classical DeePC and MPC. Therefore, in this general scenario, we linearize the LPV system at a nominal operating point $\theta_{nom}$. As the data-driven component of SD-MPC relies on the difference between the actual system and the parametric model, we assume an inaccurate parametric model by linearizing at a different operating point $\theta_{nom,M}$. Given that data-driven predictive control methods are predestined for complex and large-scale systems, the deviation of the estimated parametric model from the real system would be a natural occurrence in practical applications, thereby justifying the assumption sufficiently. The nominal operating points $\theta_{nom}$ and $\theta_{nom,M}$ for linearizing the given LPV system in \eqref{eq:LPV system} can be determined using \eqref{eq:theta} and the chosen values for the nominal states $x_{nom,1}=x_{nom,2}=15 \hspace{.15cm}[\text{cm}]$ and $x_{nom,M,1}=x_{nom,M,2}=10 \hspace{.15cm}[\text{cm}]$.

The sampling frequency for discretizing the LTI systems was chosen to be sufficiently high, equaling ten times the frequency of the fastest pole of the system linearized at the nominal operating point $\theta_{nom}$. This sufficiently large factor guarantees that the Nyquist-Shannon sampling theorem is satisfied across all the linearization points, thus ensuring accurate capture of the system dynamics. The resulting sampling time is $T_s = 2.69\text{s}$.
The input constraint set $\mathcal{U} = { u \in \mathbb{R} \mid 0 \leq u \leq 22 \hspace{.15cm}[\text{V}]}$ is bounded within the operating voltage range of the pump system, as described in \cite{Teppa_2025}. 
The output constraint set $\mathcal{Y} = { y \in \mathbb{R} \mid 0 \leq y \leq 100 \hspace{.15cm} [\text{cm}]}$  represents the allowable range of water levels in Tank 2. 
The constraint set $\mathcal{X} = { x \in \mathbb{R}^2 \mid 0 \leq x \leq 100 \hspace{.15cm} [\text{cm}]}$ can be introduced additionally to also limit the water level in Tank 1. 
However, it is important to note that this state constraint set $\mathcal{X}$ is only available for SD-MPC and not for DeePC, and is only an approximation of the real state, as much as the parametric model is only an estimate of the real system.  

\begin{figure}[]  
    \centering
    \includegraphics[width=.98\linewidth]{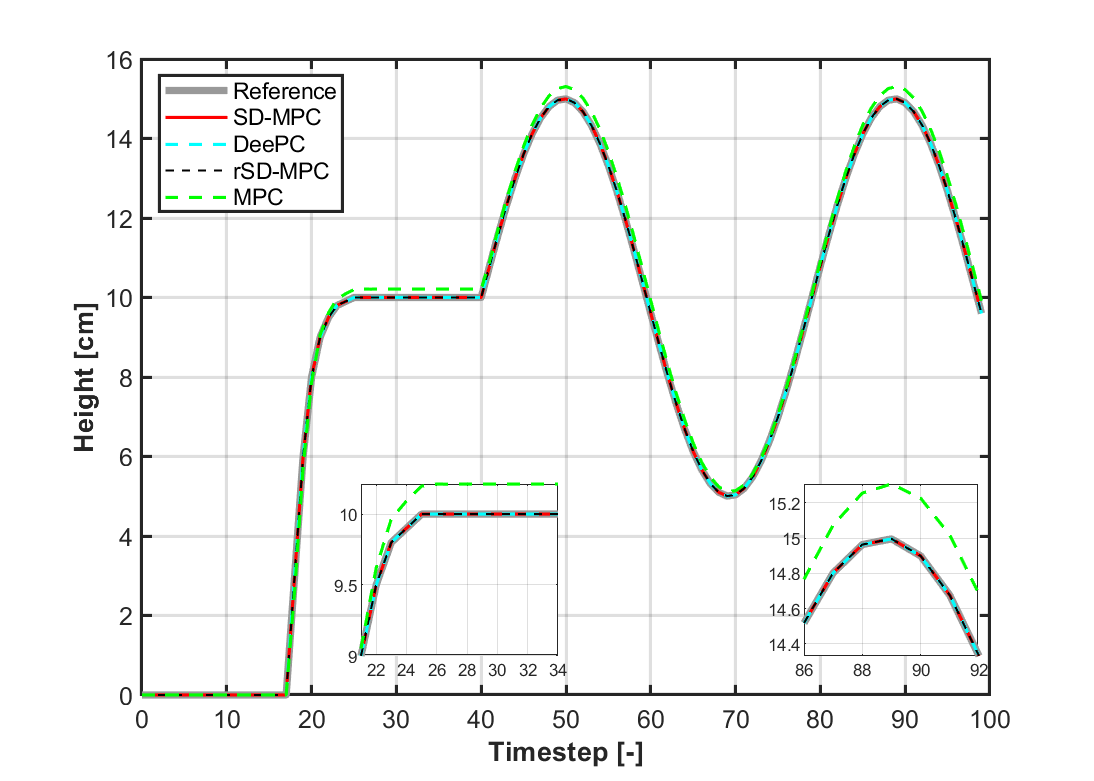}  
    \caption{General scenario: Response of the linearized system at $x_{nom}$ controlled by SD-MPC, rSD-MPC, MPC, and DeePC to follow the reference.}
    \label{fig:General_response}   
\end{figure}

Figure \ref{fig:General_response} presents the closed-loop performance of SD-MPC, rSD-MPC, MPC, and DeePC, including a response to a smoothened step reference followed by a sinusoidal reference trajectory. SD-MPC and rSD-MPC demonstrate excellent trajectory tracking performance, identical to the responses observed with the DeePC frameworks. In all three control approaches, no notable steady-state or transient error can be observed. MPC, on the other hand, exhibits sub-optimal control performance as it relies on the linearized model at $\theta_{nom,M}$, which significantly deviates from the actual LTI-system linearized at $\theta_{nom}$. In contrast, if the parametric model accurately represents the system, MPC achieves, self-evidently, the same optimal control performance as SD-MPC, rSD-MPC, and DeePC. These results show that even with an inaccurate estimated parametric model, SD-MPC can achieve optimal control performance due to the contribution of the residual data-driven component, which compensates for these inaccuracies. 
To further investigate SD-MPC and rSD-MPC, we included inaccuracies in the parametric model parameters $B_M$ and $C_M$, in addition to the previously discussed linearization of $A_M$ at the operating point $\theta_{nom,M}$. The findings demonstrated unchanged closed-loop performance, as the residual data-driven component is able to capture those discrepancies in the data collection process.

\begin{figure}[]  
    \centering
    \includegraphics[width=.98\linewidth]{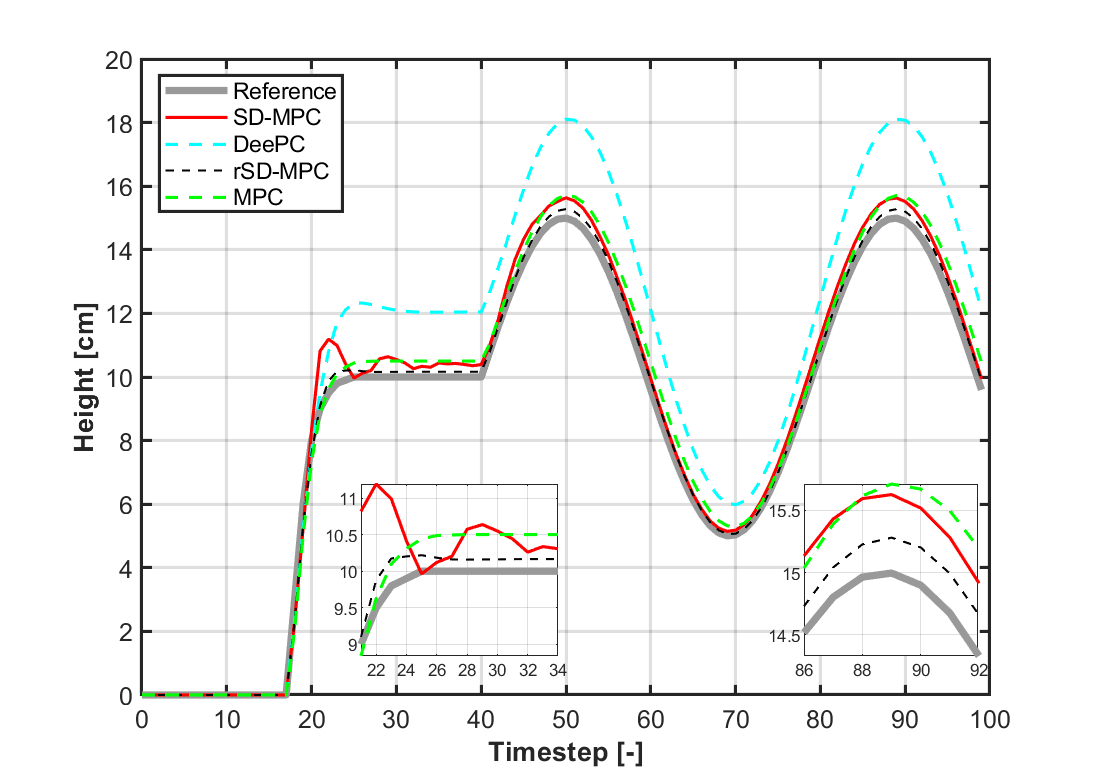}  
    \caption{Robust scenario: Response of the linearized system at $x_{nom,R}$ controlled by SD-MPC, rSD-MPC, MPC, and DeePC to follow the reference.}
    \label{fig:Robust_response}   
\end{figure}

To assess the improved robustness of the semi-data-driven approaches compared to standard DeePC within the LTI system framework, we linearize the given plant at a different operating point, $\theta_{nom,R}$, with $x_{nom,R,1}=x_{nom,R,2}=30 \hspace{.15cm}[\text{cm}]$. This setup introduces a significant mismatch between the plant model used for data collection and the plant model employed during operation. This robust scenario is intended to simulate the previously mentioned unseen operating points, where the collected data trajectories no longer precisely represent the actual system dynamics.
Figure \ref{fig:Robust_response} presents the closed-loop performance of SD-MPC, rSD-MPC, MPC, and DeePC for the robust scenario tracking the same reference trajectory. The results demonstrate the improved robustness of rSD-MPC over general SD-MPC over standard DeePC. 
More precisely, rSD-MPC and SD-MPC have a significantly lower steady-state error, and the transient response exhibits the smallest initial overshoot for rSD-MPC.
This indicates that for these unseen operating points, both SD-MPC approaches demonstrate an adaptation towards the more robust performance characteristics of classical MPC as shown in Figure~\ref{fig:Robust_response}. 

\subsection{LPV system}
\begin{figure}
    \centering
    \includegraphics[width=.98\linewidth]{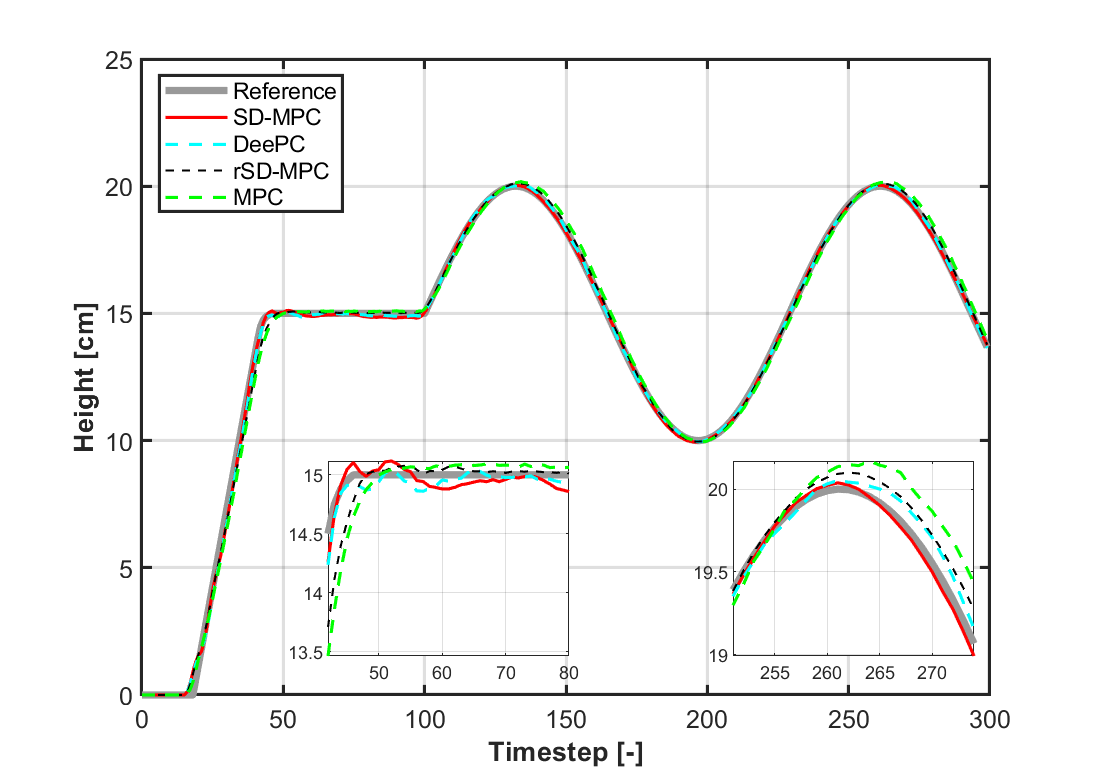}
    \caption{General scenario: Response of the nonlinear system controlled by SD-MPC, rSD-MPC, MPC, and DeePC.}
    \label{fig:LPV_Response_General}
\end{figure}

Having demonstrated the general performance and robustness of SD-MPC within the LTI system framework, we now investigate its applicability to a nonlinear system. To this end, we utilize the same nonlinear cascaded two-tank system as the LPV description presented in \eqref{eq:LPV system} and \eqref{eq:theta}. Data collection was carried out following the procedure outlined in Section~\ref{section:Semi}, which covers the range of water level in tank 2 of $y = [10,20] \hspace{.15cm} [\text{cm}]$. The parametric model used in SD-MPC is linearized around the operating point $x_1 = 5 \hspace{.15cm} [\text{cm}]$ and $x_2=15 \hspace{.15cm} [\text{cm}]$ and the previously introduced constraint sets $\mathcal{U}, \mathcal{X}, \mathcal{Y}$. The sampling time was decreased to $T_S=0.69 \text{s}$ to account for the water level-dependent dynamics. With lower water levels, the system exhibits faster poles, requiring a smaller sampling time to adequately capture the system behavior across the entire operating range. Consequently, the reference trajectory was time-scaled to accommodate this change, leading to an increased number of discrete simulation steps.
Figure~\ref{fig:LPV_Response_General} illustrates the closed-loop performance of SD-MPC, rSD-MPC, MPC, and DeePC when tracking the reference trajectory within the range of the collected data, comparable to the general scenario. All approaches demonstrate, in general, adequate control performance, with the SD-MPC method having the lowest root-mean-square error (RMSE) according to a numerical assessment. In contrast, MPC shows the poorest control behavior, which can be attributed to its sole reliance on the linearized parametric model.

\begin{figure}
    \centering
    \includegraphics[width=.98\linewidth]{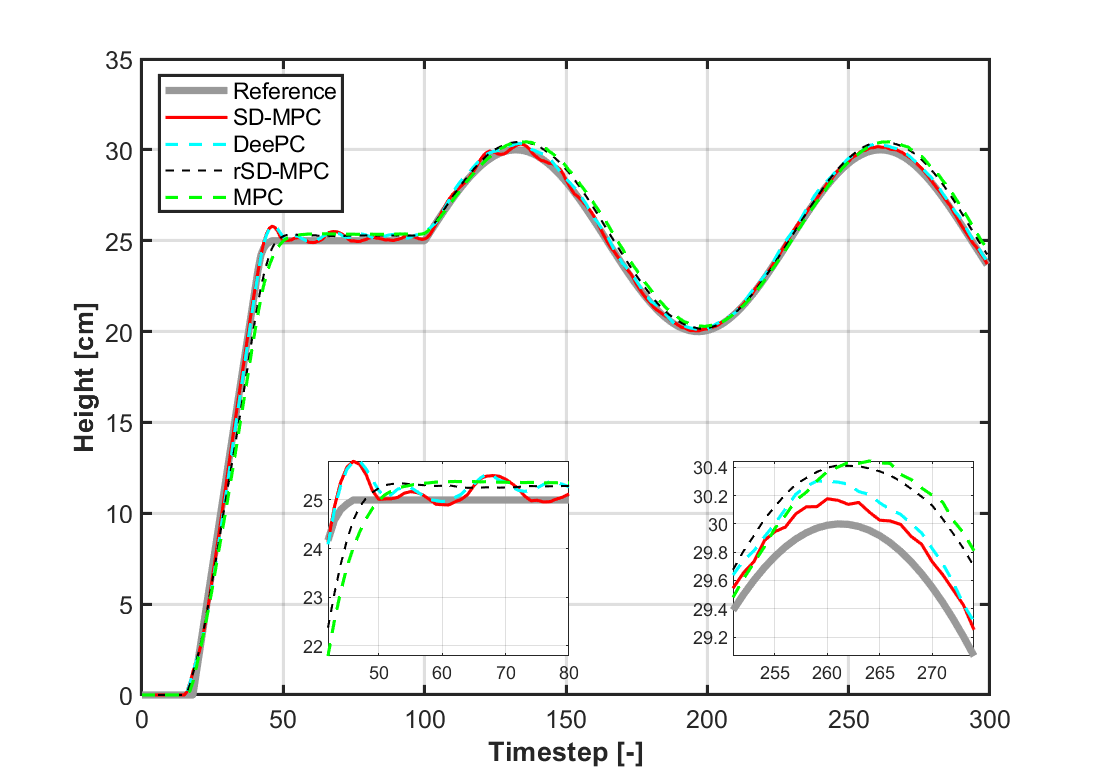}
    \caption{Robust scenario: Response of the nonlinear system controlled by SD-MPC, rSD-MPC, MPC, and DeePC.}
    \label{fig:LPV_Response_Robust}
\end{figure}

Figure~\ref{fig:LPV_Response_Robust} shows the performance of SD-MPC, rSD-MPC, MPC, and DeePC in the robust scenario for tracking a data trajectory shifted to a region beyond the original data collection range, introduced as unseen operating points.
The findings indicate that all methods can effectively control the nonlinear systems, exhibiting a robust performance with a modest deviation from the reference trajectory. Furthermore, the transient behavior suggests a closer similarity between SD-MPC and DeePC, as well as between rSD-MPC and MPC. This can be attributed to the updating rule of the parametric model's state, as explained previously in Section~\ref{section:Semi}, where the constant updating is tying rSD-MPC closer to its parametric model and therefore closer to MPC. The numerical analysis indicates that SD-MPC achieved the lowest RMSE compared to the other control approaches evaluated. In contrast, the MPC and rSD-MPC exhibit poorer performance, as shown in the magnified region of Figure~\ref{fig:LPV_Response_Robust} on the right.
The suboptimal performance of MPC can be attributed to its sole reliance on the linearized parametric model. In the case of rSD-MPC, this can be explained by the fact that only the residual data-driven component is regularized with a slack variable, meaning that only this component is robustified against nonlinearities, while the fundamental parametric model remains purely linear. The constant updating of the parametric model's state in rSD-MPC results in it relying more heavily on the linearized behavior of classical MPC at each time step, resulting in worse performance compared to SD-MPC.
It is anticipated that by developing a nonlinear SD-MPC approach, the robustness to unseen operating points can also be increased for nonlinear systems, comparable to the results observed for the LTI-system framework.

\subsection{Hyper-Parameter Optimization}
A hyperparameter optimization was performed for the DeePC, MPC, SD-MPC, and rSD-MPC control methods, including both the general scenario and the robust scenario for the LTI and LPV framework. For the evaluation of the hyperparameter tuning, the RMSE between the reference trajectory and the process output was used as the objective function. Table \ref{tab:hyperparameter} presents the optimal hyperparameter configurations resulting from the tuning process, which were also utilized for the previous experimental results. The Hankel matrix was constructed using a fixed number of $T_d=200$ data points. The output cost matrix $Q = diag(q) \in \mathbb{R}^{n_y \times n_y}$ and the input cost matrix $R =diag(r)\in \mathbb{R}^{n_u \times n_u}$ were both defined as diagonal matrices.

\begin{table}[]
\caption{Optimal hyperparameter configurations.}
\centering
\begin{tabular}{l|ll|lll}
                & \multicolumn{2}{c|}{\textbf{LTI system}}                                                          & \multicolumn{3}{c}{\textbf{LPV system}}                                                                                  \\ \cline{2-6} 
                & \multicolumn{1}{l|}{\begin{tabular}[c]{@{}l@{}}SD-MPC\\ DeePC\\ rSD-MPC\end{tabular}} & MPC       & \multicolumn{1}{l|}{\begin{tabular}[c]{@{}l@{}}SD-MPC\\ DeePC\end{tabular}} & \multicolumn{1}{l|}{rSD-MPC}   & MPC       \\ \hline
$q$             & \multicolumn{1}{l|}{$10^{4}$}                                                            & $10^{4}$     & \multicolumn{1}{l|}{$10^{4}$}                                                  & \multicolumn{1}{l|}{$10^{4}$}     & $10^{4}$     \\ \hline
$r$             & \multicolumn{1}{l|}{$10^{-3}$}                                                        & $10^{-3}$ & \multicolumn{1}{l|}{$10^{-2}$}                                              & \multicolumn{1}{l|}{$10^{-2}$} & $10^{-2}$ \\ \hline
$T_{fut}$       & \multicolumn{1}{l|}{5}                                                                & 5         & \multicolumn{1}{l|}{5}                                                      & \multicolumn{1}{l|}{5}         & 5         \\ \hline
$T_{ini}$       & \multicolumn{1}{l|}{5}                                                                & -         & \multicolumn{1}{l|}{20}                                                     & \multicolumn{1}{l|}{15}        & -         \\ \hline
$\lambda_{ini}$ & \multicolumn{1}{l|}{$10^{7}$}                                                         & -         & \multicolumn{1}{l|}{$10^{7}$}                                               & \multicolumn{1}{l|}{$10^{7}$}  & -         \\ \hline
$\lambda_{g}$   & \multicolumn{1}{l|}{1}                                                                & -         & \multicolumn{1}{l|}{$10^{4}$}                                               & \multicolumn{1}{l|}{$10^{4}$}  & -        
\end{tabular}
\label{tab:hyperparameter}
\end{table}

\section{Conclusion}
\label{section:Conclusion}
We introduced SD-MPC, which integrates physical insights into DeePC. This hybrid framework utilizes a parametric model to capture the fundamental system behavior, while the data-driven component represents the residuals. SD-MPC leverages the ease of use of DeePC while improving robustness to unseen operating points. The numerical results demonstrate that both SD-MPC and rSD-MPC exhibit comparable performance to DeePC in tracking reference trajectories within the collected data range. However, when dealing with more challenging scenarios, such as operating in regions beyond the original data collection range, the SD-MPC frameworks outperform standard DeePC in the LTI system framework in terms of robustness and trajectory tracking accuracy due to its MPC characteristics. In the LPV system framework, which involved a nonlinear system, the semi-data-driven approaches had no significantly improved robustness compared to DeePC and MPC, due to the underlying parametric LTI model. However, the results showed that both SD-MPC and rSD-MPC were still able to effectively control the nonlinear system.

Future work focuses on improving the computational complexity of data-driven control approaches with SD-MPC and exploring the potential of using lower-quality data in the data collection process. Additionally, we aim to apply SD-MPC to larger and more complex systems, including the integration of model reduction techniques. Furthermore, we intend to develop an adaptive SD-MPC approach and investigate the applicability of the SD-MPC framework for nonlinear systems.

\bibliographystyle{IEEEtran}  
\bibliography{sources}  

\end{document}